\documentclass[journal]{IEEEtran}
\usepackage{epsfig,rotating,setspace,latexsym,amsmath,epsf,amssymb,bm,theorem}
\usepackage{cite,psfrag}
\usepackage{amsfonts,authblk,subfigure}
\usepackage{graphicx}
\usepackage{color,algorithm,algpseudocode}

\newcommand{\dout}[1]{\mathcal{O}_d({#1})}

\newcommand{\mye}[2]{E_{{#1}{#2}}}
\newcommand{\bpar}[2]{\frac{\partial {#1}}{\partial {#2}}}

\newcommand{\blambda}{\bar{\lambda}}

\newcommand{\bpow}{b_{li}}
\newcommand{\ben}{G_{ni}}
\newcommand{\bsum}{s_i}
\newcommand{\bsumv}{\mathbf{s}}

\newtheorem{lemma}{Lemma}

\newtheorem{remark}{Remark}

\newenvironment{Proof}[1]{\medskip\par\noindent{\bf Proof:\,}\,#1}{{\mbox{\,$\blacksquare$}\medskip\par}}
\allowdisplaybreaks

\begin{document}

\title{Optimal Energy and Data Routing in Networks \\ with Energy Cooperation}

\author{Berk Gurakan,~\IEEEmembership{Student Member,~IEEE}, Omur Ozel,~\IEEEmembership{Member,~IEEE} and Sennur Ulukus,~\IEEEmembership{Member,~IEEE}\thanks{Manuscript received March 23, 2015; revised July 30, 2015; accepted September 13, 2015. The editor coordinating the review of this paper and approving it for
publication was W. Zhang.}
\thanks{Berk Gurakan and Sennur Ulukus are with the Department of Electrical and Computer Engineering, University of Maryland, College Park, MD 20742. Emails: \{gurakan, ulukus\}@umd.edu.}
\thanks{Omur Ozel was with the Department of Electrical and Computer Engineering, University of Maryland, College Park, MD, 20742. He is now with the Department of Electrical Engineering and Computer Sciences, University of California, Berkeley, CA, 94720. Email: ozel@berkeley.edu.}
\thanks{This work was supported by NSF Grants CNS 13-14733, CCF 14-22111 and CCF 14-22129, and presented in part at the IEEE Asilomar Conference, Pacific Grove, CA, November 2014.}}

\maketitle
\vspace{-1.0cm}

\begin{abstract}
We consider the delay minimization problem in an energy harvesting communication network with energy cooperation. In this network, nodes harvest energy from nature to sustain the power needed for data transmission, and may transfer a portion of their harvested energies to neighboring nodes through energy cooperation. For fixed data and energy routing topologies, we determine the optimum data rates, transmit powers and energy transfers, subject to flow and energy conservation constraints, in order to minimize the network delay. We start with a simplified problem where data flows are fixed and optimize energy management at each node for the case of a single energy harvest per node. This is tantamount to distributing each node's available energy over its outgoing data links and energy transfers to neighboring nodes. For this case, with no energy cooperation, we show that each node should allocate more power to links with more noise and/or more data flow. In addition, when there is energy cooperation, our numerical results indicate that, energy is routed from nodes with lower data loads to nodes with higher data loads. We then extend this setting to the case of multiple energy harvests per node over time. In this case, we optimize each node's energy management over its outgoing data links and its energy transfers to neighboring nodes, over multiple time slots. For this case, with no energy cooperation, we show that, for any given node, the sum of powers on the outgoing links over time is equal to the single-link optimal power over time. Finally, we consider the problem of joint flow control and energy management for the entire network. We determine the necessary conditions for joint optimality of a power control, energy transfer and routing policy. We provide an iterative algorithm that updates the data flows, energy flows and power distribution over outgoing data links sequentially. We show that this algorithm converges to a Pareto-optimal operating point.\\ \\
\end{abstract}

\begin{keywords}
Energy cooperation, energy harvesting, wireless energy transfer, optimal routing, resource allocation
\end{keywords}

\section{Introduction}

\begin{figure}[t]
\begin{center}
\includegraphics[scale=0.26]{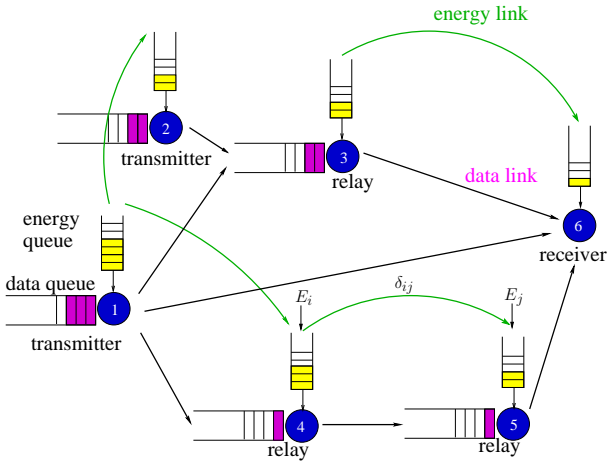}
\end{center}
\caption{System model.}
\label{sysmod1}
\vspace{-0.5cm}
\end{figure}

We consider an energy harvesting communication network with energy cooperation as shown in Fig.~\ref{sysmod1}. Each node harvests energy from nature and all nodes may share a portion of their harvested energies with neighboring nodes through energy cooperation \cite{gurakan_subm}. We focus on the delay minimization problem for this network. The delay on each link depends on the information carrying capacity of the link, and in particular, it decreases monotonically with the capacity of the link for a fixed data flow through it; see e.g., \cite[eqn.~(5.30)]{Bertsekas92}. The capacity, in turn, is a function of the power allocated to the link, and in particular, it is a monotonically increasing function of the power, for instance, through a logarithmic Shannon type capacity-power relationship; see e.g., \cite[eqns.~(9.60) and (9.62)]{cover_book}. In addition, the delay on a link is a monotonically increasing function of the data flow through it, for a fixed link capacity \cite[eqn.~(5.30)]{Bertsekas92}.

In this paper, we consider the joint data routing and capacity assignment problem for this setting under fixed data and energy routing topologies \cite[Section~5.4.2]{Bertsekas92}. Our work is related to and builds upon classical and recent works on data routing and capacity assignment in communication networks \cite{Bertsekas92, gallager1977minimum, bertsekas1984second, gavish1989system, bertsekas1998network, yen2001near, cruz2003optimal, xiao2004simultaneous,  cui2007cross, xi2008node}, and recent works on energy harvesting communications \cite{jingtcom12, kayatcom12, ozel11, ho-zhang-tsp12, finite} and energy cooperation \cite{gurakan_subm, kaya13ita, kaya13itw, varshney2008, grover2010, doost, lshi, zhang2013mimo, zhou2012wireless, simeonegraph, huang2014, gurakanisit2014, ding2015application, hu2013utility, guo2014joint, leithon2014energy, chia2014energy, xu2015cost} in wireless networks. In our previous work \cite{gurakan_subm, gurakanisit2014}, we studied the optimal energy management problem for several basic multi-user network structures with energy harvesting transmitters and one-way wireless energy transfer. Inspired by joint routing and resource allocation problems in the classical works such as \cite{gallager1977minimum, bertsekas1984second, gavish1989system, bertsekas1998network, xiao2004simultaneous,xi2008node}, in our current work, we study joint routing of energy and data in a general multi-user scenario with data and energy transfer. We specialize in the objective of minimizing the total delay in the system. To the best of our knowledge, this problem has not been addressed in the context of energy harvesting wireless networks with energy cooperation. Among previous works, the approach that is most related to ours is that in reference \cite{simeonegraph}, which studies networkwide optimization of energy and information flows in communication networks with simultaneous energy and information transfer. We also note the references \cite{doost,lshi} for related joint data routing and energy transfer schemes in networks with special energy transfer capabilities and no energy harvesting. Finally, we refer the reader to \cite{hu2013utility, guo2014joint, leithon2014energy, chia2014energy, xu2015cost} for a related line of research about resource allocation in base stations powered by renewable energy  and energy cooperation.

We divide our development in this paper into three parts. In the first part, we assume that the data flows through the links are fixed, and each node harvests energy only once. In this setting, we determine the optimum energies allocated to outgoing data links of the nodes and the optimum amounts of energies transferred between the nodes. In the second part, we extend this setting to the case of multiple energy harvests for each node. In the last part, we optimize both data flows on the links and energy management at the nodes. We determine the jointly optimal routing of data and energy in the network as well as distribution of power over the outgoing data links at each node. 

In the first part of the paper, in Section~\ref{singleEH}, we focus on the optimal energy management problem at the nodes with a single energy harvest at each node. First, we consider the case without energy cooperation. We show that this problem can be decomposed into individual problems, each one to be solved for a single node. We show that more power should be allocated to links with more noise and/or more data flow, resembling channel inversion type of power control \cite{goldsmith-varaiya}. Next, we consider the case with energy cooperation, where nodes transfer a portion of their own energies to neighboring nodes. In this case, we have the joint problem of energy routing among the network nodes and energy allocation among the outgoing data links at each node. For this problem, we develop an iterative algorithm that visits all energy links sufficiently many times and decreases the network delay monotonically. We numerically observe that energy flows from nodes with lightly loaded data links to nodes with heavily loaded data links.

In the second part of the paper, in Section~\ref{multiEH}, we extend our setting to the case of multiple energy harvests at each node, by allowing time-varying energy harvesting rates over large time frames. We incorporate the time variation in the energy harvests and solve for the optimal energy management at each node and energy routing among the nodes. First, we focus on the case without energy cooperation. We show that the sum powers on the outgoing data links of a node over time slots is equal to the single-link optimal transmit power of that node over time and can be found using \cite{jingtcom12, kayatcom12, ozel11}. When the optimal sum powers are known, we show that the problem reduces to a problem with a single energy arrival and can be solved using our method. Next, we focus on the case with energy cooperation. We show that this problem can be mapped to the original problem with no energy cooperation by constructing an equivalent directed graph.

In the last part of the paper, in Section~\ref{joint}, we consider the problem of determining the jointly optimal data and energy flows in the network and the power distribution over the outgoing data links at all nodes. We determine a set of necessary conditions for the joint optimality of a power control, energy transfer and data routing policy. We then develop an iterative algorithm that updates the data flows, energy flows and distribution of power over the outgoing data links at each node in a sequential manner. We show that this algorithm converges to a Pareto-optimal operating point.

\section{Network Flow and Energy Model}

We use directed graphs to represent the network topology, and data and energy flows through the network. All nodes are energy harvesting, and are equipped with separate wireless energy transfer units. Information and energy transfer channels are orthogonal to each other.

\subsection{Network Data Topology}

We represent the data topology of the network by a directed graph. In this model, a collection of nodes, labeled $n=1,\dots,N$, can send and receive data across communication links.In particular, a node can be either a source node, a destination node or a relay node. A data communication link is represented as an ordered pair $(i,j)$ of distinct nodes. The presence of a link $(i,j)$ means that the network is able to send data from the start node $i$ to the end node $j$. We label the data links as $l=1,\dots,L$. The network data topology can be represented by an $N \times L$ matrix, $\mathbf{A}$, in which every entry $A_{nl}$ is associated with node $n$ and link $l$ via
\begin{align}
A_{nl} =
\begin{cases}
1,  &\text{if $n$ is the start node of data link $l$} \\
{-}1, &\text{if $n$ is the end node of data link $l$} \\
0, &\text{otherwise}
\end{cases}
\end{align}
We define $\mathcal{O}_d (n)$ as the set of outgoing data links from node $n$, and $\mathcal{I}_d(n)$ as the set of incoming data links to node $n$. We define $N$-dimensional vector $\mathbf{s}$ whose $n$th entry $s_n$ denotes the non-negative amount of exogenous data flow injected into the network at node $n$. On each data link $l$, we let $t_l$ denote the amount of flow and we call the $L$-dimensional vector $\mathbf{t}$ the flow vector. At each node $n$, the flow conservation implies:
\begin{align}
\sum_{l \in \mathcal{O}_d(n)} t_l - \sum_{l \in \mathcal{I}_d(n)} t_l = s_n, \quad \forall n
\end{align}
The flow conservation law over all the network can be compactly written as:
\begin{align}
\mathbf{A} \mathbf{t} = \mathbf{s}
\end{align}
We define $c_l$ as the information carrying capacity of link $l$. Then, we require $t_l \leq c_l,\, \forall l$.

\subsection{Network Energy Topology}

All nodes are equipped with energy harvesting units. In this section, we describe the energy model for the case of a single energy harvest per node. We present the extension to the case of multiple energy harvests in Section~\ref{multiEH}. Here, each node $n$ harvests energy in the amount of $E_n$. We use $N$-dimensional vector $\mathbf{E}$ to denote the energy arrival vector for the system. In the energy cooperation setting, there are energy links similar to data links. An energy link is represented as an ordered pair $(i,j)$ of distinct nodes where the presence of an energy link means that it is possible to send energy from the start node to the end node. Energy links are labeled as $q=1,\dots,Q$. Energy transfer efficiency on each energy link is denoted with $0 < \alpha_q \leq 1$ which means that when $\delta$ amount of energy is transferred on link $q$ from node $i$ to node $j$, node $j$ receives $\alpha_q \delta$ amount of energy. We assume that the directionality and the position of energy transfer links are fixed whereas the amount of energy transferred on these links are unknown. The network energy topology can be represented by an $N \times Q$ matrix, $\mathbf{B}$, in which every entry $B_{nq}$ is associated with node $n$ and energy link $q$ via
\begin{align}
B_{nq} =
\begin{cases}
1, & \text{if $n$ is the start node of energy link $q$} \\
-\alpha_q, &\text{if $n$ is the end node of energy link $q$} \\
0, &\text{otherwise}
\end{cases}
\end{align}
On each energy link $q$, we let $y_q$ be the amount of energy transferred. We call the $L$-dimensional vector $\mathbf{y}$ the energy flow vector. We denote by $\mathcal{O}_e (n)$ and $\mathcal{I}_e(n)$, respectively, the sets of outgoing and incoming energy links at node $n$.

\subsection{Communication Model and Delay Assumptions}

Following the M/M/1 queueing model in \cite{Bertsekas92}, we represent the delay on data link $l$ as:
\begin{align}\label{delay_exp}
D_{l}  = \frac{t_{l}}{c_{l} - t_{l}}
\end{align}
where $t_l$ is the flow and $c_l$ is the information carrying capacity of link $l$, with $t_l \leq c_l,\, \forall l$. This delay expression is a good approximation for systems with Poisson arrivals at the entry points, exponential packet lengths and moderate-to-heavy traffic loads \cite{Bertsekas92}. In view of energy scarcity in the network, moderate-to-heavy traffic load assumption generally holds. The packet arrival and packet length assumptions are made for convenience of analysis. Moreover, we assume that the slot length is sufficiently large to enable convergence to stationary distributions. In particular, we assume that the slot length is sufficiently longer than the average delay yielded by the M/M/1 approximation. Each node $n$, on the transmitting edge of data link $l$, with channel noise $\sigma_l$, enables a capacity $c_l$ by expanding power $p_{l}$. These quantities are related by the Shannon formula \cite[eqn.~(9.60)]{cover_book} as:
\begin{align}
c_l = \frac{1}{2} \log \left(1 + \frac{p_l}{\sigma_l} \right)  \label{cappower}
\end{align}
where all $\log$s in this paper are with respect to base $e$. At each node $n$, the total power expanded on data and energy links are constrained by the available energy, i.e.,
\begin{align}
\sum_{l \in \mathcal{O}_d(n)} p_l +  \sum_{q \in \mathcal{O}_e(n)} y_q \leq E_n +  \sum_{q \in \mathcal{I}_e(n)} \alpha_q y_q, \quad \forall n
\end{align}
Using $L$-dimensional vector $\mathbf{p} = (p_1, \dots, p_L)$ and $\mathbf{F} = \mathbf{A}^+$ where $(A^+)_{nl} = \max \{A_{nl},0\}$, the energy availability constraints can be compactly written as:
\begin{align}
\mathbf{Fp} + \mathbf{By} \leq \mathbf{E}
\label{powercon}
\end{align}
We note that we use power and energy interchangeably in (\ref{powercon}) and in the rest of the paper by assuming slot lengths of 1 unit.

\section{Capacity Assignment Problem for Single Time Slot} \label{singleEH}

In this section, we consider the capacity assignment problem for the case of a single energy harvest per node. We assume that the flow assignments, $t_l$, on all links are fixed and are serviceable by the harvested energies and energy transfers. The total delay in the network is:
\begin{align}
 D = \sum_l \frac{t_l}{c_l - t_l}
\end{align}
The capacity assignment problem, with the goal of minimizing the total delay in the network is:
\begin{align}
\min_{c_l, p_l, y_q} \qquad & \sum_l \frac{t_l}{c_l - t_l}  \nonumber \\
\mbox{s.t.} \qquad & \mathbf{Fp} + \mathbf{By} \leq \mathbf{E} \nonumber \\
& t_l \leq c_l, \quad \forall l
\end{align}
By using the capacities $c_l$ in (\ref{cappower}), we write the problem in terms of the link powers $p_l$ and energy transfers $y_q$ only as:
\begin{align}
\min_{p_l, y_q} \qquad & \sum_l \frac{t_l}{\frac{1}{2} \log \left(1 + \frac{p_l}{\sigma_l} \right) - t_l} \nonumber  \\
\mbox{s.t.} \qquad & \mathbf{Fp} + \mathbf{By} \leq \mathbf{E} \nonumber \\
& p_l \geq \sigma_l \left(e^{2 t_l} -1\right), \quad \forall l \label{problem1m}
\end{align}
We solve the problem in (\ref{problem1m}) in the rest of this section. We first identify some structural properties of the optimal solution in the next sub-section. The following analysis relies on the standing assumption that this problem has at least one feasible solution. To see if this problem is feasible, one can replace the objective function of (\ref{problem1m}) with a constant and solve a feasibility problem, which turns out to be a linear program.

\subsection{Properties of the Optimal Solution}

First, we note that the objective function can be written in the form $\sum_i f_i(g(x_i))$ where $f_i(x_i) = \frac{t_i}{x_i-t_i}$ and $g(x_i) = \frac{1}{2} \log{(1+x_i)}$. Since $f$ is convex and non-increasing and $g$ is concave, the resulting composition function is convex \cite{boyd}. The constraint set is affine. Therefore, (\ref{problem1m}) is a convex optimization problem. The Lagrangian function is:
\begin{align}
\mathcal{L} =  & \sum_l \frac{t_l}{\frac{1}{2} \log\left(1 + \frac{p_l}{\sigma_l}\right) - t_l} + \sum_n \lambda_n \left[ \sum_{l \in \mathcal{O}_d(n)} p_l \right. \nonumber \\
& + \left. \sum_{q \in \mathcal{O}_e(n)} y_q - E_n -  \sum_{q \in \mathcal{I}_e(n)} \alpha_q y_q \right] \nonumber  \\
& - \sum_l \beta_l \left[ p_l - \sigma_l \left(e^{2 t_l} -1\right) \right] - \sum_q \theta_q y_q
\end{align}
where $\lambda_n$ and $\beta_l$ are Lagrange multipliers corresponding to the energy constraints of the nodes in (\ref{powercon}) and the feasibility constraints $t_l \leq c_l$, respectively. The KKT optimality conditions are:
\begin{align}
h_l' (p_l) + \lambda_{n(l)} - \beta_l &= 0, \quad \forall l \label{KKT1} \\
\lambda_{m(q)} - \alpha_q \lambda_{k(q)} - \theta_q &= 0, \quad \forall q \label{KKT2}
\end{align}
where $h_l (p_l) \triangleq t_l \left(\frac{1}{2} \log{ \left(1 + \frac{p_l}{\sigma_l} \right)} - t_l \right)^{-1}$, $n(l)$ is the beginning node of data link $l$, $m(q)$ and $k(q)$ are the beginning and end nodes of energy link $q$, respectively. The additional complementary slackness conditions are: 
\begin{align}
& \lambda_n \left( \sum_{l \in \mathcal{O}_d(n)} p_l \right. +   \sum_{q \in \mathcal{O}_e(n)} y_q - E_n -  \left. \sum_{q \in \mathcal{I}_e(n)} \alpha_q y_q \right) = 0, \, \forall n \label{slack1} \\
& \beta_l \left[ p_l - \sigma_l \left(e^{2 t_l} -1\right) \right] = 0, \quad \forall l \label{slack2} \\
& \theta_q y_q =0, \quad \forall q \label{slack3}
\end{align}

We now identify some properties of the optimal power allocation in the following three lemmas.

\begin{lemma}
If the problem in (\ref{problem1m}) is feasible, then $\beta_l = 0, \, \forall l$.
\label{lemma1}
\end{lemma}

\begin{Proof}
If the problem in (\ref{problem1m}) is feasible, its objective function must be bounded. Equality in the second set of constraints in (\ref{problem1m}) for any $l$ implies that the objective function is unbounded. Therefore, we must have strict inequality in those constraints for all $l$, and from (\ref{slack2}), we conclude that $\beta_l =0, \forall l$.
\end{Proof}

\begin{lemma}
\label{lemma2}
At every node $n$, the optimal power allocation amongst outgoing data links satisfies
\begin{align}
h_l'(p_l) = h_m'(p_m), \quad \forall l,m \in \mathcal{O}_d(n)
\end{align}
\end{lemma}

\begin{Proof}
From (\ref{KKT1}) and Lemma~\ref{lemma1} we have,
\begin{align}
h_l' (p_l) = - \lambda_{n(l)}, \quad \forall l
\end{align}
For outgoing data links $l$ and $m$ that belong to the same node $n$,
\begin{align}
h_l' (p_l) = - \lambda_{n} = h_m' (p_m)
\end{align}
which gives the desired result.
\end{Proof}

\begin{lemma}
\label{lemma3}
If some energy is transferred through energy link $q$ across nodes $(i,j)$, then,
\begin{align}
h_l'(p_l) = \alpha_q h_m'(p_m), \quad \forall l \in \mathcal{O}_d(i), \, \forall m \in \mathcal{O}_d(j)
\end{align}
\end{lemma}

\begin{Proof}
If some energy is transferred through energy link $q$, then $y_q > 0$, and from (\ref{slack3}), $\theta_q = 0$. From (\ref{KKT2}), we have,
\begin{align}
\lambda_i = \alpha_q \lambda_j \label{lemma3proof1}
\end{align}
Writing (\ref{KKT1}) for nodes $i$ and $j$, we have,
\begin{align}
h_l'(p_l) &= - \lambda_i, \quad \forall l \in \mathcal{O}_d(i) \label{lemma3proof2} \\
h_m'(p_m) &= - \lambda_j, \quad \forall m \in \mathcal{O}_d(j) \label{lemma3proof3}
\end{align}
and the result follows from combining  (\ref{lemma3proof1}), (\ref{lemma3proof2}) and (\ref{lemma3proof3}).
\end{Proof}

In the following two sub-sections, we separately solve the problem for the cases of no energy transfer and with energy transfer.

\subsection{Solution for the Case of No Energy Transfer} \label{notransfersol}

In the case of no energy transfer, we have $y_q = 0, \, \forall q$, and the problem becomes only in terms of $p_l$ as stated below:
\begin{align}
\min_{p_l} \qquad & \sum_l \frac{t_l}{\frac{1}{2} \log \left(1 + \frac{p_l}{\sigma_l} \right) - t_l} \nonumber  \\
\mbox{s.t.} \qquad & \sum_{l \in \mathcal{O}_d(n)} p_l \leq E_n, \quad \forall n \nonumber \\
& p_l \geq \sigma_l \left(e^{2 t_l} -1\right), \quad \forall l
\end{align}
This problem can be decomposed into $N$ sub-problems as:
\begin{align}
\min_{p_l} \qquad & \sum_n \sum_{l \in \mathcal{O}_d(n)} \frac{t_l}{\frac{1}{2} \log{ \left(1 + \frac{p_l}{\sigma_l} \right)} - t_l}  \nonumber \\
\mbox{s.t.} \qquad &  \sum_{l \in \mathcal{O}_d(n)} p_l \leq E_n, \quad \forall n  \nonumber \\
& p_l \geq \sigma_l (e^{2 t_l} -1), \quad \forall l
\end{align}
Since the constraint set depends only on the powers of node $n$, there is no interaction between the nodes. Every node will independently solve the following optimization problem:
\begin{align}
\min_{p_l} \qquad & \sum_{l \in \mathcal{O}_d(n)} \frac{t_l}{\frac{1}{2} \log\left(1 + \frac{p_l}{\sigma_l}\right) - t_l} \nonumber \\
\mbox{s.t.} \qquad & \sum_{l \in \mathcal{O}_d(n)} p_l \leq E_n \nonumber \\
& p_l \geq \sigma_l \left(e^{2 t_l} -1\right), \quad \forall l \in \mathcal{O}_d(n) \label{nodeproblem}
\end{align}
The feasibility of (\ref{nodeproblem}) requires $E_n \geq \sum_{l \in \mathcal{O}_d(n)} \sigma_l (e^{2 t_l} -1)$ which we assume holds. Similar to (\ref{problem1m}), (\ref{nodeproblem}) is a convex optimization problem with the KKT optimality conditions:
\begin{align}
h_l' (p_l) + \lambda &= 0, \quad \forall l \in \dout{n} \label{KKT1new}
\end{align}
with the complementary slackness condition:
\begin{align}
\lambda \left( \sum_{l \in \mathcal{O}_d(n)} p_l  - E_n \right) &= 0 \label{slack1new}
\end{align}
The Lagrange multipliers for the second set of constraints in (\ref{nodeproblem}) are not included, because similar to Lemma~\ref{lemma1}, they will always be satisfied with strict inequality. From (\ref{KKT1new}), we have
\begin{align}
- \lambda &= h_l'(p_l) \\
&= \frac{-t_l}{2 \sigma_l} \left[ \frac{1}{2} \log{ \left(1 + \frac{p_l}{\sigma_l} \right)} - t_l \right]^{-2} \left( 1 + \frac{p_l}{\sigma_l} \right)^{-1} \label{optpeq}
\end{align}
After some algebraic manipulations shown in Appendix~\ref{appendixlambert}, we have
\begin{align}
p_l (\lambda) = \sigma_l \left( e^{2( W(z_l) + t_l )} - 1 \right) \label{optp}
\end{align}
where $z_l = \sqrt{\frac{t_l e^{-2 t_l}}{2 \lambda \sigma_l}}$ and $W(\cdot)$ is the Lambert W function defined as the inverse of the function $w \rightarrow w e^w$ \cite{corless1996lambertw}. Next, we prove some monotonicity properties for the optimal solution, as a function of the qualities of the channels and the amounts of data flows through the channels.

\begin{lemma}
For fixed $t_l$, $p_l$ is monotone increasing in $\sigma_l$.
\end{lemma}

\begin{Proof}
By differentiating (\ref{optp}) and using the following property \cite{corless1996lambertw}
\begin{align}
\frac{dW(x)}{dx} = \frac{W(x)}{x(1 + W(x))} \label{lambertprop}
\end{align}
it can be verified as shown in Appendix \ref{appendixdiffpower} that
\begin{align}
\frac{\partial p_l}{\partial \sigma_l} &= e^{2 t_l} \frac{e^{2 W(z_l)}}{1 + W(z_l)} -1 >0
\label{diffpower}
\end{align}
where the inequality follows from $e^{2 t_l} > 1,\, \forall t_l >0$, and $\frac{e^{2 z}}{1+z} > 1, \, \forall z>0$, proving the lemma.
\end{Proof}

This lemma shows that, for fixed data flows, more power should be allocated to channels with more noise power, similar to channel inversion power control \cite{goldsmith-varaiya}.

\begin{lemma}
For fixed $\sigma_l$, $p_l$ is monotone increasing in $t_l$.
\end{lemma}

\begin{Proof}
By differentiating (\ref{optp}), it can be verified as shown in Appendix~\ref{appendixdiffflow} that
\begin{align}
\frac{\partial p_l}{\partial t_l} &= \frac{\sigma_l ( W(z_l) + 2 t_l ) \, e^{2(W(z_l) + t_l)}}{t_l (1+W(z_l))}  > 0
\label{diffflow}
\end{align}
proving the lemma.
\end{Proof}

This lemma shows that, for fixed channel qualities (i.e., fixed noise powers), more power should be allocated to links with more data flow.

Finally, we solve (\ref{nodeproblem}) as follows: From the total energy constraint, we have $\sum_l p_l (\lambda^*) = E_n$. We perform a one dimensional search on $\lambda$ to find $\lambda^*$ that satisfies $\sum_l p_l (\lambda^*) = E_n$, where $p_l (\lambda^*)$ is given in (\ref{optp}). Once $\lambda^*$ is obtained, the optimal power allocations are found from (\ref{optp}).

\subsection{Solution for the Case with Energy Transfer} \label{transfersol}

Now, we consider the case with energy transfer, i.e., $y_q \geq 0$ for some $q$. Assume that some energy $y_q>0$ is transferred from node $i$ to node $j$ on energy link $q$. Writing (\ref{optp}) for the outgoing data links of node $i$ and node $j$, we have,
\begin{align}
p_l (\lambda_i) &= \sigma_l \left( e^{2( W(z_{il}) + t_l )} - 1 \right), \quad \forall l \in \mathcal{O}_d(i) \label{pnodei} \\
p_l (\lambda_j) &= \sigma_l \left( e^{2( W(z_{jl}) + t_l )} - 1 \right), \quad \forall l \in \mathcal{O}_d(j) \label{pnodej}
\end{align}
where $z_{il} = \sqrt{\frac{t_l e^{-2 t_l}}{2 \lambda_i \sigma_l}}$ and $z_{jl} = \sqrt{\frac{t_l e^{-2 t_l}}{2 \lambda_j \sigma_l}}$. From (\ref{lemma3proof1}), we have $\lambda_i = \alpha_q \lambda_j$. The energy causality constraints on node $i$ and node $j$ are:
\begin{align}
\sum_{l \in \mathcal{O}_d(i)} p_l (\lambda_i^*) &=  E_i - y_q \label{energycausi} \\
\sum_{l \in \mathcal{O}_d(j)} p_l (\lambda_j^*) &= E_j + \alpha_q y_q \label{energycausj}
\end{align}
Equations (\ref{lemma3proof1}), (\ref{energycausi}) and (\ref{energycausj}) imply
\begin{align}
\alpha_q \sum_{l \in \mathcal{O}_d(i)} p_l (\alpha_q \lambda_j^*) + \sum_{l \in \mathcal{O}_d(j)} p_l (\lambda_j^*) = \alpha_q E_i + E_j \label{energyequal}
\end{align}
which can be solved by a one-dimensional search on $\lambda_j^*$.

\begin{algorithm}
\caption{Algorithm to solve capacity assignment problem for single time slot}
\label{algo1}
\begin{algorithmic}[1]
\Statex \textbf{Initialize} \Comment{No energy transfer}
\Statex \vspace{-0.1in} \hrulefill
\For{$i = 1:N$}
	 \State Find $\lambda_i$ such that $\sum_{l \in \mathcal{O}_d(i)} p_l(\lambda_i) = E_i$, $p_l$ is (\ref{optp}) 	
\EndFor
\Statex \vspace{-0.1in} \hrulefill
\Statex \textbf{Main Algorithm}
\Statex \vspace{-0.1in} \hrulefill
\For{$q=1:Q$} \Comment{All energy links}
\State Set $(i,j) \leftarrow$ (origin,destination) of energy link $q$
\If{$\lambda_i < \alpha_q \lambda_j$} \Comment{Perform energy transfer} \newline Find $\lambda_j^*$ such that \newline $\alpha_q \sum_{l \in \mathcal{O}_d(i)} p_l (\alpha_q \lambda_j^*) + \sum_{l \in \mathcal{O}_d(j)} p_l (\lambda_j^*) = \alpha_q E_i + E_j$ \newline
Set $\text{Tap}_q =  E_i - \sum_{l \in \mathcal{O}_d(i)} p_l (\alpha_q \lambda_j^*)$ \Comment{Update tap level}\newline
\Comment{Update battery levels} \newline Set $E_i = \sum_{l \in \mathcal{O}_d(i)} p_l (\alpha_q \lambda_j^*), E_j = \sum_{l \in \mathcal{O}_d(j)} p_l (\lambda_j^*)$ 
\ElsIf{$\lambda_i > \alpha_q \lambda_j$} \Comment{Recall some energy} 
\While{$\text{Tap}_q \geq 0, \lambda_i > \alpha_q \lambda_j, E_j \geq 0$} \newline
\Comment{Recall $\epsilon$ energy} \newline Set $E_i = E_i + \epsilon, E_j = E_j - \alpha_q \epsilon, \text{Tap}_q = \text{Tap}_q - \epsilon$ \newline 
Find $\lambda_i, \lambda_j$ such that \newline $E_i = \sum_{l \in \mathcal{O}_d(i)} p_l (\lambda_i), E_j = \sum_{l \in \mathcal{O}_d(j)} p_l (\lambda_j)$
\EndWhile
\EndIf
\EndFor
\end{algorithmic}
\end{algorithm}

We solve (\ref{problem1m}) by iteratively allowing energy to flow through a single link at a time provided all links are visited infinitely often. Since we do not know which energy links will be active in the optimal solution, we may need to call back any transferred energy in the previous iterations. To perform this, we keep track of transferred energy over each energy link by means of meters as in \cite{gurakan_subm}. Initially, we start from the no energy transfer solution and compute $\lambda_n$ for every node $n$ as described in the previous section. At every iteration, we open only one energy link $q$ at a time, and whenever energy flows through link $q$, (\ref{energyequal}) must be satisfied with $E_i$ and $E_j$ in (\ref{energyequal}) replaced with the battery levels of nodes $i$ and $j$ at the current iteration. In particular, if $\lambda_i < \alpha_q \lambda_j$, we search for $\lambda_j^*$ that satisfies (\ref{energyequal}). If no solution to (\ref{energyequal}) can be found, this means $\lambda_i > \alpha_q \lambda_j$, and then previously transferred energy must be called back to the extent possible according to the meter readings. The algorithmic description is given below as Algorithm 1. From the strict convexity of the objective function, we note that each iteration decreases the objective function as described similarly in \cite[Section~V.A]{gurakan_subm}. Our algorithm converges since bounded real monotone sequences always converge, and the limit point is a local minimum because, the iterations can only stop when $\lambda_i = \alpha_q \lambda_j$ for the energy links where $y_q > 0$ which are the KKT optimality conditions from (\ref{lemma3proof1}). This local minimum is also the unique global minimum due to the convexity of the problem.

\section{Capacity Assignment Problem for Multiple Time Slots} \label{multiEH}

In this section, we consider the capacity assignment problem for the scenario where the energy arrival rates to the nodes can change over time.  We assume that the time is slotted and there are a total of $T$ equal-length slots. In slots $i=1,\dots,T$, each node $n$ harvests energy with amounts $\mye{n}{1},\mye{n}{2},\dots,\mye{n}{T}$, and the arriving energies can be saved in a battery for use in future time slots. The subscript $i$ denotes the time slot, and the quantities $t_{li}, c_{li}, p_{li}, \sigma_{li}$ and $y_{qi}$ denote the flow, capacity, power, noise power, and energy transfer in slot $i$.  We assume that the flow allocation and channel noises do not change over time, i.e., $t_{li} = t_l$ and $\sigma_{li} = \sigma_l, \forall i, \forall l$. We further assume that the slots are long enough so that the M/M/1 approximation is valid at every slot $i$. In particular, slot length is sufficiently larger than the average delay resulting from the M/M/1 approximation. Then, the average delay on link $l$ at time slot $i$ is given as,
\begin{align}
\label{delay_exptime}
D_{li}  = \frac{t_{l}}{c_{li} - t_{l}}
\end{align}
where $c_{li} = \frac{1}{2} \log{\left(1 + \frac{p_{li}}{\sigma_l} \right)}$. As the energy that has not arrived yet cannot be used for data transmission or energy transfer, the power policies of the nodes are constrained by causality of energy in time. These constraints are written as:
\begin{align}
\sum_{i=1}^k & \Bigg( \sum_{l \in \mathcal{O}_d(n)} p_{li} + \sum_{q \in \mathcal{O}_e(n)} y_{qi}  \Bigg) \nonumber \\
& \leq \sum_{i=1}^k \Bigg( E_{ni} + \sum_{q \in \mathcal{I}_e(n)} \alpha_q y_{qi} \Bigg), \quad \forall n, \, \forall k
\end{align}
The capacity assignment problem with fixed link flows to minimize the total delay over all links and all time slots can be formulated as:
\begin{align}
\min_{p_{li}, y_{qi}} \qquad & \sum_{i=1}^T \sum_l \frac{t_l}{\frac{1}{2} \log{\left(1 + \frac{p_{li}}{\sigma_l}\right)} - t_l} \nonumber \\
\mbox{s.t.} \qquad & \sum_{i=1}^k \Bigg( \sum_{l \in \mathcal{O}_d(n)} p_{li} + \sum_{q \in \mathcal{O}_e(n)} y_{qi} \Bigg) \nonumber \\
& \quad \leq \sum_{i=1}^k \Bigg( E_{ni} + \sum_{q \in \mathcal{I}_e(n)} \alpha_q y_{qi} \Bigg), \quad \forall n, \,  \forall k \nonumber \\
& p_{li} \geq \sigma_l (e^{2 t_l} -1), \quad \forall l, \, \forall i \label{problemtime}
\end{align}
The problem in (\ref{problemtime}) is convex and the Lagrangian function can be written as:
\begin{align}
\mathcal{L} = & \sum_{i=1}^T \sum_l h_l(p_{li}) + \sum_{n} \sum_{k=1}^T \lambda_{nk} \left[ \sum_{i=1}^k \left( \sum_{l \in \mathcal{O}_d(n)} p_{li} \right. \right.  \nonumber \\ 
& \left. \left. + \sum_{q \in \mathcal{O}_e(n)} y_{qi} -  E_{ni} - \sum_{q \in \mathcal{I}_e(n)} \alpha_q y_{qi} \right) \right] \nonumber\\
& - \sum_{q} \sum_{i=1}^T \theta_{qi} y_{qi}
\end{align}
where $h_l (p_{li}) \triangleq t_l \left[\frac{1}{2} \log{\left(1 + \frac{p_{li}}{\sigma_l}\right)} - t_l\right]^{-1}$. The Lagrange multipliers for the second set of constraints for (\ref{problemtime}) are not included here because similar to before, they will always be satisfied with strict inequality. The KKT optimality conditions are:
\begin{align}
h_l'(p_{li}) + \sum_{k=i}^T \lambda_{n(l)k} &= 0, \quad \forall l, \, \forall i \label{KKT1time} \\
\sum_{k=i}^T \lambda_{m(q)k} - \alpha_q \sum_{k=i}^T \lambda_{r(q)k} - \theta_{qi} &= 0, \quad \forall q, \, \forall i \label{KKT2time}
\end{align}
where $n(l)$ is the beginning node of data link $l$, $m(q)$ and $r(q)$ are the beginning and end nodes of energy link $q$. The additional complementary slackness conditions as:
\begin{align}
\lambda_{nk} \Bigg[ \sum_{i=1}^k \Bigg( \sum_{l \in \mathcal{O}_d(n)} p_{li} & + \sum_{q \in \mathcal{O}_e(n)} y_{qi} -  E_{ni} \nonumber \\ 
\quad - \sum_{q \in \mathcal{I}_e(n)} \alpha_q y_{qi} \Bigg) \Bigg] & = 0, \quad \forall n, \, \forall k \label{slack1time} \\
\theta_{qi} y_{qi} &=0, \quad \forall q, \, \forall i \label{slack2time}
\end{align}

Now, we extend Lemmas~\ref{lemma2} and \ref{lemma3} to the case of multiple energy arrivals over time.

\begin{lemma}
\label{lemmaehequal}
At every node $n$, the optimal power allocation amongst outgoing data links satisfies
\begin{align}
h_l'(p_{li}) = h_m'(p_{mi}), \quad \forall l,m \in \mathcal{O}_d(n), \, \forall i
\end{align}
\end{lemma}

\begin{Proof}
From (\ref{KKT1time}), we have,
\begin{align}
h_l' (p_{li}) = - \sum_{k=i}^T \lambda_{n(l)k}
\end{align}
For outgoing data links $l$ and $m$ that belong to the same node $n$,
\begin{align}
h_l' (p_{li}) = - \sum_{k=i}^T \lambda_{nk} = h_m' (p_{mi}), \quad \forall i
\end{align}
from which the result follows.
\end{Proof}

\begin{lemma}
\label{lemmaetequal}
If some energy is transferred through energy link $q$ across nodes $(a,b)$ at time slot $i$,
\begin{align}
h_l'(p_{li}) = \alpha_q h_m'(p_{mi}), \quad \forall l \in \mathcal{O}_d(a), \, \forall m \in \mathcal{O}_d(b)
\end{align}
\end{lemma}

\begin{Proof}
If some energy is transferred through energy link $q$ at time slot $i$, then $y_{qi} > 0$, and from (\ref{slack2time}), $\theta_{qi} = 0$. From (\ref{KKT2time}), we have,
\begin{align}
\sum_{k=i}^T \lambda_{ak} = \alpha_q \sum_{k=i}^T \lambda_{bk} \label{lemma5proof1}
\end{align}
Then, we have,
\begin{align}
h_l'(p_{li}) & = - \sum_{k=i}^T \lambda_{ak} = - \alpha_q \sum_{k=i}^T \lambda_{bk} \nonumber \\
& = \alpha_q h_m'(p_{mi}), \quad \forall l \in \mathcal{O}_d(a), \, \forall m \in \mathcal{O}_d(b) \label{lemma5proof2}
\end{align}
where the first equality follows from writing (\ref{KKT1time}) for node $a$, the second equality follows from (\ref{lemma5proof1}), and the third equality follows from writing (\ref{KKT1time}) for node $b$.
\end{Proof}

In the following two sub-sections, we separately solve the problem for the cases of no energy transfer and with energy transfer.

\subsection{Solution for the Case of No Energy Transfer}
In this case, we have $y_{qi} = 0, \forall i, \forall q$. The problem becomes only in terms of $p_{li}$ as follows:
\begin{align}
\min_{p_{li}} \qquad & \sum_{i=1}^T \sum_l \frac{t_l}{\frac{1}{2} \log{\left(1 + \frac{p_{li}}{\sigma_l}\right)} - t_l} \nonumber \\
\mbox{s.t.} \qquad & \sum_{i=1}^k \sum_{l \in \mathcal{O}_d(n)} p_{li}  \leq \sum_{i=1}^k E_{ni}, \quad \forall n, \, \forall k \nonumber \\
& p_{li} \geq \sigma_l (e^{2 t_l} -1), \quad \forall l, \, \forall i \label{problemtimenoet}
\end{align}
The problem can be decomposed into $N$ sub-problems as:
\begin{align}
\min_{p_{li}} \qquad & \sum_{i=1}^T \sum_n \sum_{l \in \mathcal{O}_d(n)} \frac{t_l}{\frac{1}{2} \log{\left(1 + \frac{p_{li}}{\sigma_l}\right)} - t_l} \nonumber \\
\mbox{s.t.} \qquad & \sum_{i=1}^k \sum_{l \in \mathcal{O}_d(n)} p_{li}  \leq \sum_{i=1}^k E_{ni}, \quad \forall n, \, \forall k \nonumber \\
& p_{li} \geq \sigma_l (e^{2 t_l} -1), \quad \forall l, \, \forall i \label{problemtimenoet2}
\end{align}
Since the constraint set depends only on the powers of node $n$, there is no interaction between the nodes. Every node will independently solve the following optimization problem:
\begin{align}
\min_{p_{li}} \qquad & \sum_{i=1}^T \sum_{l \in \mathcal{O}_d(n)} \frac{t_l}{\frac{1}{2} \log{\left(1 + \frac{p_{li}}{\sigma_l}\right)} - t_l} \nonumber \\
\mbox{s.t.} \qquad & \sum_{i=1}^k \sum_{l \in \mathcal{O}_d(n)} p_{li}  \leq \sum_{i=1}^k E_{ni}, \quad \forall k \nonumber \\
& p_{li} \geq \sigma_l (e^{2 t_l} -1), \quad \forall l \in \mathcal{O}_d(n), \, \forall i \label{problemtimenoet3}
\end{align}
Solving (\ref{problemtimenoet3}) entails finding the optimal energy management policy for each link $l$, over all time slots $i$. We define $\bpow = p_{li} - \sigma_l (e^{2 t_l} -1)$ and $\ben = E_{ni} - |\mathcal{O}_d(n)| \sigma_l (e^{2 t_l} -1)$. Then, (\ref{problemtimenoet3}) becomes:
\begin{align}
\min_{\bpow} \qquad & \sum_{i=1}^T \sum_{l \in \mathcal{O}_d(n)} \frac{t_l}{\frac{1}{2} \log{\left(e^{2 t_l} + \frac{\bpow}{\sigma_l}\right)} - t_l} \nonumber \\
\mbox{s.t.} \qquad & \sum_{i=1}^k \sum_{l \in \mathcal{O}_d(n)} \bpow  \leq \sum_{i=1}^k \ben, \quad \forall k \nonumber \\
& \bpow \geq 0, \quad \forall l \in \mathcal{O}_d(n), \, \forall i \label{problemtimenoet4}
\end{align}
For feasibility of (\ref{problemtimenoet4}) we need $\ben \geq 0$ which we assume holds. Now, we state an important property of the optimal policy which is proved in Appendix~\ref{appendixsumpow}.

\begin{lemma}
\label{lemmasumpow}
The optimal total power allocated to outgoing data links at each slot $i$, $\sum_{l \in \mathcal{O}_d(n)} \bpow$, is the same as the single-link optimal transmit power with energy arrivals $\ben$.
\end{lemma}

From Lemma~\ref{lemmasumpow} we have that the sum powers in outgoing data links are given by the single-link optimal transmit powers which can be found by the geometric method in \cite{jingtcom12} or by the directional water-filling method in \cite{ozel11}. Once the sum powers are obtained, individual link powers are found by solving $x(s_i)$ which is defined in (\ref{sumpow2}) in Appendix~\ref{appendixsumpow}. The problem in $x(s_i)$ includes a single energy harvest and is in the form of (\ref{nodeproblem}), therefore, we use the method proposed in Section~\ref{notransfersol} to find the individual link powers.

\subsection{Solution for the Case with Energy Transfer}

From (\ref{KKT1time}) and some algebraic manipulations we have
\begin{align}
p_{li} = \sigma_l \left( e^{2( W(z_{il}) + t_l )} - 1 \right) \label{pnodetime}
\end{align}
where $z_{il} = \sqrt{ \frac{t_l e^{-2 t_{l}}}{2 (\sum_{k=i}^T \lambda_{n(l)k}) \sigma_l } }$ and $W(\cdot)$ is the Lambert W function. The Lagrangian structure of this problem is more complicated compared to the previous case since the power allocation at time $i$ depends on $\{ \lambda_{n(l)k} \}_{k=i}^T$. Therefore, here, we offer an alternative solution.

In the scenario described above, the nodes have the capability to save their energies to use in future slots. We note that saving energy for use in future slots is equivalent to transferring energy to future slots with energy transfer efficiency of $\alpha = 1$. In light of this observation, an equivalent representation of (\ref{problemtime}) can be obtained by modifying the network graph where each time slot is treated as a new node with a single energy arrival and saving energy for future slots is represented by energy transfer links of efficiency 1. The modification to the network graph is performed in the following way. First, we make $T$ replicas of the network graph including all the nodes and the existing data and energy transfer links. Each replica will denote the network at one time slot. We let each replica node receive one energy harvest which amounts to the energy harvested by that node in that time slot. We keep the existing energy and data links but we add new energy links between different replicas of the same node. For every node $n$, we add energy links of efficiency $1$ between replicas $k$ and $k+1$, where $k = 1,\dots,T-1$. Relabeling the nodes, we obtain a new graph where all nodes have one energy harvest. Essentially, we have reduced this problem to the case in Section~\ref{transfersol} and we use the solution provided in that section.

We finally remark that our framework can easily be extended to address variations in channel fading coefficients and energy transfer efficiencies by allowing the noise powers $\sigma_l$ and energy transfer efficiencies $\alpha_l$ to vary from slot to slot, i.e., defining $c_{li} = \frac{1}{2} \log \left( 1 + \frac{p_{li}}{\sigma_{li}} \right)$ and replacing $\alpha_l$ with $\alpha_{li}$.

\section{Joint Capacity and Flow Optimization} \label{joint}

In this section, we consider the joint optimization of capacity and flow assignments, in contrast to capacity assignment only with fixed flows, as considered in the previous sections. We focus on the case with a single energy harvest per node as in Section~\ref{singleEH}. The delay minimization problem with joint capacity and flow allocation can be formulated as:
\begin{align}
\min_{p_l, y_q, t_l} \qquad & \sum_l \frac{t_l}{\frac{1}{2} \log{ \left( 1 + \frac{p_l}{\sigma_l} \right) } - t_l} \nonumber  \\
\mbox{s.t.} \qquad & \mathbf{Fp} + \mathbf{By} \leq \mathbf{E} \nonumber \\
& p_l \geq \sigma_l (e^{2 t_l} -1), \quad \forall l \nonumber \\
& \mathbf{At} = \mathbf{s}  \label{problemjoint}
\end{align}
where we optimize not only the powers $p_l$ and energy transfers $y_q$, but also the data flows $t_l$. In (\ref{problemjoint}), the first set of constraints are the energy constraints, the second set of constraints are the capacity constraints on individual links, and the last set of constraints are the flow conservation constraints at all nodes.

We assume that the exogenous arrivals $\mathbf{s}$ is serviceable by the energy harvests and energy transfers. This means that problem (\ref{problemjoint}) has a bounded solution and furthermore no data link is operating at the capacity, i.e., the capacity constraints are never satisfied with equality unless $t_l = p_l = 0$. We solve the problem in (\ref{problemjoint}) in the remainder of this section. Here, the constraint set is convex, however, the objective function is not jointly convex in $p_l$ and $t_l$ \cite{Bertsekas92}, therefore, (\ref{problemjoint}) is not a convex optimization problem. We study the necessary optimality conditions by writing the Lagrangian function as follows:
\begin{align}
\mathcal{L}  = & \sum_l \frac{t_l}{\frac{1}{2} \log{ \left(1 + \frac{p_l}{\sigma_l} \right)} - t_l} + \sum_n \lambda_n \Bigg[ \sum_{l \in \mathcal{O}_d(n)} p_l \nonumber \\
& + \sum_{q \in \mathcal{O}_e(n)} y_q - E_n -  \sum_{q \in \mathcal{I}_e(n)} \alpha_q y_q \Bigg] - \sum_l \beta_l [p_l \nonumber \\
& - \sigma_l (e^{2 t_l} -1) ] + \sum_n \nu_n \Bigg[ \sum_{l \in \mathcal{O}_d(n)} t_l - \sum_{l \in \mathcal{I}_d(n)} t_l - s_n \Bigg] \nonumber \\
& - \sum_q \theta_q y_q - \sum_l \gamma_l t_l \label{lagjoint}
\end{align}
The KKT optimality conditions are:\footnote{With the objective function of (\ref{problemjoint}), there is an uncertainty when $t_l = p_l = 0$. Nonetheless, we argue as in \cite[page~441]{Bertsekas92} that the objective function of (\ref{problemjoint}) is differentiable over the set of all $p_l$ with $\frac{1}{2} \log{ \left(1 + \frac{p_l}{\sigma_l} \right)} > t_l$ and $\bpar{\mathcal{L}}{p_l} = 0, \bpar{\mathcal{L}}{t_l} = 0$ and $\bpar{\mathcal{L}}{y_q} = 0$ are necessary conditions for optimality.}
\begin{align}
& \frac{-t_l}{2 \sigma_l} \left[ \frac{1}{2} \log{ \left(1 + \frac{p_l}{\sigma_l} \right)} - t_l  \right]^{-2} \left( 1 + \frac{p_l}{\sigma_l} \right)^{-1} + \lambda_{n(l)} \nonumber \\
& \qquad - \beta_l  = 0, \quad \forall l \label{jointKKT1} \\
& \frac{1}{2} \log{\left( 1 + \frac{p_l}{\sigma_l} \right)} \left[ \frac{1}{2} \log{ \left(1 + \frac{p_l}{\sigma_l} \right)} - t_l  \right]^{-2} \nonumber \\
& \qquad + \nu_{n(l)} - \nu_{m(l)} - \gamma_l + 2 \beta_l \sigma_l e^{2 t_l}  = 0, \quad \forall l \label{jointKKT2} \\
& \lambda_{k(q)} - \alpha_q \lambda_{z(q)} - \theta_q = 0, \quad \forall q \label{jointKKT3}
\end{align}
where $n(l)$ and $m(l)$ are the source and destination nodes of data link $l$, $k(q)$ and $z(q)$ are the source and destination nodes of energy link $q$, respectively. The complementary slackness conditions are:
\begin{align}
& \lambda_n \left( \sum_{l \in \mathcal{O}_d(n)} p_l +  \sum_{q \in \mathcal{O}_e(n)} y_q - E_n -  \sum_{q \in \mathcal{I}_e(n)} \alpha_q y_q \right) = 0, \, \forall n \label{jointslack1} \\
& \nu_n \left( \sum_{l \in \mathcal{O}_d(n)} t_l - \sum_{l \in \mathcal{I}_d(n)} t_l - s_n \right) = 0, \quad \forall n \label{jointslack5} \\
& \theta_q y_q  = \gamma_l t_l  = 0, \quad \forall q, \, \forall l \label{jointslack2} \\
& \beta_l \left[ p_l - \sigma_l (e^{2 t_l} -1) \right]  = 0, \quad \forall l \label{jointslack4} \\
& \lambda_n, \beta_l, \theta_q, \gamma_l \geq 0,\quad \forall l, \, \forall q, \, \forall n \label{jointslack6} 
\end{align}
We note that $\nu_n < 0$ is allowed since the Lagrange multiplier $\nu$ corresponds to an equality constraint. Lemma \ref{jointlemma}, proved in Appendix~\ref{appendixjoint}, states the necessary optimality conditions.

\begin{lemma}
\label{jointlemma}
For a feasible set of flow variables $\{t_l\}_{l=1}^L$, transmission power allocations $\{p_l\}_{l=1}^L$ and energy transfers $\{y_q\}_{q=1}^Q$ to be the solution to the problem in (\ref{problemjoint}), the following conditions are necessary.\\
1) For every node $n$, there exists a constant $\lambda_n >0$ such that
\begin{align}
\frac{t_l}{2 \sigma_l} \left[ \frac{1}{2} \log{ \left(1 + \frac{p_l}{\sigma_l} \right)} - t_l  \right]^{-2} & \left( 1 + \frac{p_l}{\sigma_l} \right)^{-1} \leq \lambda_n, \nonumber \\ 
& \quad \forall l \in \mathcal{O}_d(n) \label{ncon1}
\end{align}
and with equality if $p_l > 0$.\\
2) For every node $n$, there exists a constant $\tilde{\nu}_n \geq 0$ such that
\begin{align}
\sum_{l \in \mathcal{F}_{n,d}} \frac{1}{2} \log{\left(1 + \frac{p_l}{\sigma_l} \right)} & \left[ \frac{1}{2} \log{ \left(1 + \frac{p_l}{\sigma_l} \right)} - t_l  \right]^{-2}  = \tilde{\nu}_n, \nonumber \\
& \quad \forall d = 1,\dots,D \label{ncon2}
\end{align}
where $\mathcal{F}_{n,d}$ is a data path that starts from node $n$ and ends at destination node $d$ and for which $p_l > 0, \forall l \in \mathcal{F}_{n,d}$. The condition in (\ref{ncon2}) is valid for all such data paths that start from node $n$ and end at any destination node.\\
3) For all energy transfer links $q$, and $\forall l \in \mathcal{O}_d(n), \forall k \in \mathcal{O}_d(m)$ such that $p_l > 0$ and $p_k > 0$ where $n$ and $m$ are the origin and destination nodes of energy transfer link $q$
\begin{align}
\frac{t_l}{2 \sigma_l} & \left[ \frac{1}{2} \log{ \left(1 + \frac{p_l}{\sigma_l} \right)} - t_l  \right]^{-2}  \left( 1 + \frac{p_l}{\sigma_l} \right)^{-1} \geq \nonumber \\
& \alpha_q \frac{t_k}{2 \sigma_k} \left[ \frac{1}{2} \log{ \left(1 + \frac{p_k}{\sigma_k} \right)} - t_k  \right]^{-2} \left( 1 + \frac{p_k}{\sigma_k} \right)^{-1} \label{ncon3}
\end{align}
where (\ref{ncon3}) is satisfied with equality if $y_q > 0$.
\end{lemma}

From Lemma~\ref{jointlemma}, the structure of the optimal solution is as follows: We define $h_l(p_l,t_l)$ as the objective function of the problem in  (\ref{problemjoint}), $h_l(p_l,t_l) \triangleq t_l\left[{\frac{1}{2} \log{\left(1 + \frac{p_l}{\sigma_l} \right)} - t_l } \right]^{-1}$. We see from (\ref{ncon1}) that nodes should allocate more power on links where the quantity $\left|\bpar{h_l}{p_l}\right|$ is large and less power on links where this quantity is small. Similarly, from (\ref{ncon2}), we see that less flow should be allocated on paths where the quantity $\sum\limits_{l \in_{\mathcal{F}_{n,d}}} \bpar{h_l}{t_l}$ is large and more flow on paths where this quantity is small. Finally, (\ref{ncon3}) tells us the necessary conditions for energy transfer. We describe our solution to the problem in (\ref{problemjoint}) in the next section.

\subsection{Algorithmic Solution for the Joint Capacity and Flow Optimization Problem}

In this section, we propose an iterative algorithm. There are three steps to each iteration as summarized below. We start from a feasible point $(\mathbf{t}_0, \mathbf{p}_0)$.

\begin{enumerate}
\item {\it Energy Management Step:} We fix a stepsize $\xi_p > 0$. Each node computes $\bpar{h_l}{p_l}$ for their own outgoing data links where $p_l > 0$. Every node performs the following iteration:
\begin{align}
\label{energymanage}
p_l^{k+1} =
\begin{cases} p_l^k + \xi_p, & \mbox{if } l = \arg \max_{l \in \mathcal{O}_d(n)} \left|\bpar{h_l}{p_l}\right|  \\
p_l^k - \xi_p, & \mbox{if } l = \arg \min_{l \in \mathcal{O}_d(n)} \left|\bpar{h_l}{p_l}\right|  \\
p_l^k, & \mbox{otherwise }
\end{cases}
\end{align}
where $k$ denotes the iteration number, and the derivatives are computed at the current iteration, i.e., for $(\mathbf{t}^k,\mathbf{p}^k)$.

\item {\it Data Routing Step:} We fix a stepsize $\xi_t > 0$. Each node $n$ computes $\sum\limits_{l \in{\mathcal{F}_{n,d}}} \bpar{h_l}{t_l}$ for the data paths originating from source node $n$ and ending at any destination. Assume the path ${\mathcal{F}^*_{n}}$ maximizes $\sum\limits_{l \in{\mathcal{F}_{n,d}}} \bpar{h_l}{t_l}$ and the path ${\mathcal{G}^*_{n}}$ minimizes $\sum\limits_{l \in{\mathcal{F}_{n,d}}} \bpar{h_l}{t_l}$ for each $n$. Every node performs the following iteration:
\begin{align}
\label{datarouting}
t_l^{k+1} =
\begin{cases} t_l^k - \xi_t, & \mbox{if } l \in {\mathcal{F}^*_{n}}  \\
t_l^k + \xi_t, & \mbox{if } l \in {\mathcal{G}^*_{n}}  \\
t_l^k, & \mbox{otherwise }
\end{cases}
\end{align}

\item {\it Energy Routing Step:} This step is the same as described in Section~\ref{transfersol}. Specifically, every node goes through its energy transfer links and makes the comparison $\left|\bpar{h_l}{p_l}\right| \gtrless \alpha_q \left|\bpar{h_m}{p_m}\right|$ where $m$ is the receiving node of energy link $q$. If $\left|\bpar{h_l}{p_l}\right| < \alpha_q \left|\bpar{h_m}{p_m}\right|$, then some energy is transferred through link $q$. If $\left|\bpar{h_l}{p_l}\right| > \alpha_q \left|\bpar{h_m}{p_m}\right|$, then some energy must be called back, as explained in Section \ref{transfersol}.

\item Go back to step 1, or terminate if sufficiently many iterations are performed.
\end{enumerate}

\begin{algorithm}
\caption{Algorithm to solve joint capacity and flow assignment problem for single time slot}
\label{algo2}
\begin{algorithmic}[1]
\Statex \textbf{Initialize}
\Statex \vspace{-0.1in} \hrulefill
\State Generate initial point
\Statex \vspace{-0.1in} \hrulefill
\Statex \textbf{Energy management step}
\Statex \vspace{-0.1in} \hrulefill
\For{$n=1:N$} \Comment{All nodes} \newline
Find $\arg \max_{l \in \mathcal{O}_d(n)} \bpar{h_l}{p_l}$, perform (\ref{energymanage}) as long as $p_l \geq \sigma_l (e^{2 t_l} -1)$ is still satisfied 
\EndFor
\Statex \vspace{-0.1in} \hrulefill
\Statex \textbf{Data routing step}
\Statex \vspace{-0.1in} \hrulefill
\For{$n=1:N$} \Comment{All Nodes} \newline
Find path $\mathcal{F}_n^*$ that maximizes and ${\mathcal{G}^*_{n}}$ that minimizes $\sum\limits_{l \in{\mathcal{F}_{n,d}}} \bpar{h_l}{t_l}$ where $d \in \mathcal{O}_d(n)$ 
\For{$l \in \mathcal{F}_n^*$} $t_l^{k+1} = t_l^k - \xi_t$
\EndFor
\For{$l \in \mathcal{G}_n^*$} $t_l^{k+1} = t_l^k + \xi_t$  as long as $p_l \geq \sigma_l (e^{2 t_l} -1)$ is still satisfied 
\EndFor
\EndFor
\Statex \vspace{-0.1in} \hrulefill
\Statex \textbf{Energy routing step}
\Statex \vspace{-0.1in} \hrulefill
\For{$q=1:Q$} \Comment{All energy links}
\State Set $(i,j) \leftarrow$ (origin,destination) of energy link $q$
\State Set $\lambda_i = \left| \bpar{h_i}{p_i} \right|$ and $\lambda_j = \bpar{h_j}{p_j}$
\State Use steps $6:10$ of Algorithm 1
\EndFor
\Statex \vspace{-0.1in} \hrulefill
\State Repeat until convergence
\end{algorithmic}
\end{algorithm}

We describe our Algorithm in tabular form as Algorithm 2 below. We note that our algorithm reduces to the one in \cite{xi2008node} in the case of no energy harvesting or energy transfer. Next, we discuss the convergence and optimality properties of our algorithm.

\subsection{Convergence and Optimality Properties of the Proposed Algorithm}

Every iteration of the algorithm decreases the objective function and the iterations are bounded. Using the fact that real monotone bounded sequences converge, we conclude that the algorithm converges. Assume $(\mathbf{t}^*,\mathbf{p}^*,\mathbf{y}^*)$ is a convergence point of the algorithm. Next, we show that this point satisfies the KKT optimality conditions stated in Lemma~\ref{jointlemma}.

\begin{lemma}
$(\mathbf{t}^*,\mathbf{p}^*,\mathbf{y}^*)$ satisfies the conditions stated in Lemma~\ref{jointlemma}.
\end{lemma}
\begin{Proof}
When the algorithm converges, we must have $p_l^{k+1} = p_l^k$. From (\ref{energymanage}), this is only possible when $\bpar{h_l}{p_l}$ is constant for $l \in \mathcal{O}_d(n)$ which is equivalent to (\ref{ncon1}). Similarly, we must have $t_l^{k+1} = t_l^k$ and from (\ref{datarouting}), this is only possible when $\sum\limits_{l \in{\mathcal{F}_{n,d}}} \bpar{h_l}{t_l}$ is constant over all paths, which is equivalent to (\ref{ncon2}). Using a similar argument we conclude that energy transfers satisfy (\ref{ncon3}). This means that $(\mathbf{t}^*,\mathbf{p}^*,\mathbf{y}^*)$ satisfies Lemma \ref{jointlemma}.
\end{Proof}

Now, we remark that even though we cannot claim global optimality of the solution, we have the following {\it Pareto-optimality} condition.

\begin{remark}
\label{pareto}
Assume that $(\mathbf{t}^*,\mathbf{p}^*,\mathbf{y}^*)$ satisfies the conditions stated in Lemma~\ref{jointlemma}, then the vector of link delays is Pareto-optimal, i.e., there does not exist another pair of feasible allocations $(\hat{\mathbf{t}}, \hat{\mathbf{p}},\hat{\mathbf{y}})$ such that
\begin{align}
h_l(\hat{p}_l,\hat{t}_l) \leq h_l(p_l^*,t_l^*), \quad \forall l
\end{align}
with at least one inequality being strict.
\end{remark}

This remark means that at the Pareto-optimal point, the average delay cannot be strictly reduced on one link without it being increased on another. The proof of this remark follows similar lines as the proof in \cite[Thm.~4]{xi2008node} and is omitted here for brevity. We note that, in particular, any local optimal point is Pareto-optimal due to the fact that local optimal points satisfy KKT conditions in Lemma \ref{jointlemma}.

\section{Numerical Results}

In this section, we give simple numerical results to illustrate the resulting optimal policies. We study three network topologies shown in Figs.~\ref{sysmod2}, \ref{sysmod3} and \ref{sysmod4}. For all examples, we assume $\sigma_l = 0.1$ units $\forall l$. The slot length is of 1 unit for convenience, so that we use power and energy; rate and data interchangeably.

\subsection{Network Topology 1}
\begin{figure}[t]
\begin{center}
\includegraphics[scale=0.29]{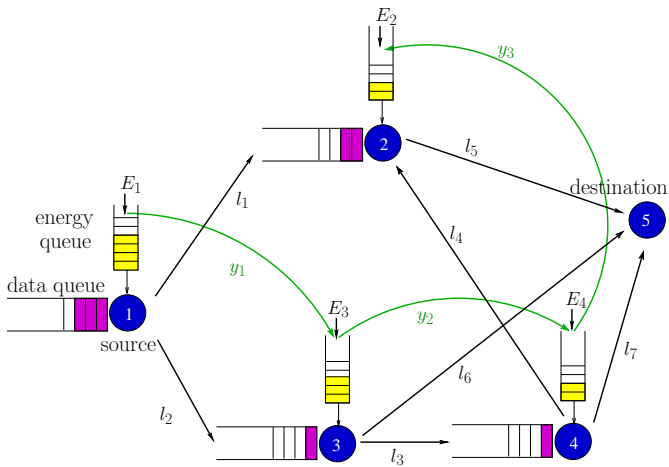}
\end{center}
\vspace{0.5cm}
\caption{Network topology 1.}
\label{sysmod2}
\end{figure}

We first consider the network topology in Fig.~\ref{sysmod2} with one source, one destination and three relays in between. The data and energy links are shown and labeled as in Fig.~\ref{sysmod2}, where $l_i$s represent data links and $y_q$s represent energy links. The fixed data flows are $\mathbf{t} = [t_1, \dots, t_7] = [2, 1, 0.5, 0.125, 2.125, 0.375, 0.5]$ units. We consider two time slots. The energy arrival vector is $\mathbf{E} = [(E_{11},E_{12}), \dots, (E_{41}, E_{42})] = [(15,10),(8,6),(5,9),(1,6)]$ units and energy transfer efficiencies are $\boldsymbol{\alpha} = [\alpha_1, \alpha_2, \alpha_3] = [0.6, 0.5, 0.5]$.

The optimal energy transfer vector is found as $\mathbf{y} = [(y_{11},y_{12}),(y_{21},y_{22}),(y_{31},y_{32})] = [(0,3.75), \break (3.93, 9.52), (2.35,9.81)]$ units and power allocation vector after energy transfer is $\mathbf{p} = [(p_{11},p_{12}),\dots, (p_{71},p_{72})] = [\,(7.5, 7.5), \,(3.13, 3.13), \,(0.62, 1), \,(0.13, 0.22), \,(9.17, 11), \break (0.45, 0.74), \, (0.48, 0.73)]$ units. Lemmas~\ref{lemmaehequal} and \ref{lemmaetequal} can be verified numerically: $h_l'(p_{li})$ equalizes for different outgoing links of the same node, for example, on links $l_1$ and $l_2$ (Lemma~\ref{lemmaehequal}); and where some energy is transferred, $h_l'(p_{li})$ is proportional to the energy transfer efficiency of that energy transfer link, for example, $h_2'(p_{22})/h_3'(p_{32})=\alpha_1$ (Lemma~\ref{lemmaetequal}). Lemma~\ref{lemmasumpow} can also be verified numerically: after the energy transfers, the sum powers of the links are the optimal single-link powers. For example, node 1 has harvested $(15,10)$ energies and transferred $(0,3.75)$ of them. Equivalently node 1 has harvested $(15,6.25)$ and the single-link optimal powers for these harvests are $(10.625, 10.625)$ which is $(p_{11}+p_{21}, p_{12}+p_{22})$. It is interesting to note that node 4 has transferred more energy than it initially had, which means that most of the transferred energy has been routed from other nodes. This is due to the high data flow on link $l_5$ which leads to a higher energy demand at node 2.

\subsection{Network Topology 2}

\begin{figure}[t]
\begin{center}
\includegraphics[scale=0.28]{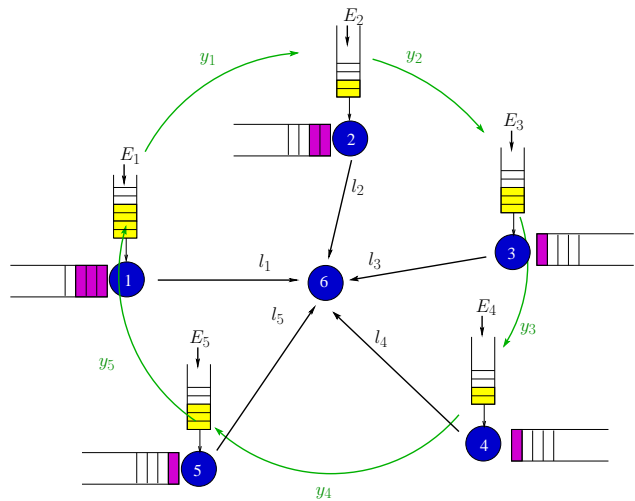}
\end{center}
\caption{Network topology 2.}
\label{sysmod3}
\end{figure}

We next consider the star topology in Fig.~\ref{sysmod3} where five sources are communicating with one destination similar to a multiple access scenario. The data flows are $\mathbf{t} = [0.5, 2, 0.5, 0.5, 2]$ units. We consider a single time slot. The energy arrivals to all the nodes are the same, i.e., $E_n = 15$ units, $\forall n$. The wireless energy transfer efficiencies are $\alpha_q = 0.5, \forall q$.

The optimal energy transfer vector is found as $\mathbf{y} = [11.92, 0, 9.66, 16.29, 0]$ units and the power vector after energy transfer is $\mathbf{p} = [3.07, 20.96, 5.33, 3.53, 23.15]$ units. This system is symmetric in terms of energy arrivals, channel noises and energy transfer efficiencies, and furthermore $t_1 = t_3 = t_4$ and $t_2 = t_5$. In this scenario, one might expect $p_1 = p_3 = p_4$ and $p_2 = p_5$. However, in the optimal solution $p_5 > p_2$. The reason for this asymmetry is as follows. Due to the high data loads on links $l_2$ and $l_5$, there is no incentive for these nodes to share their energy. Then, in the optimal solution, $y_2 = y_5 = 0$ and nodes $2$ and $5$ act as energy sink nodes where energy is collected and not sent out. We see that node 5 has two nodes transferring energy to it while node 2 has only one node transferring energy. Then, $p_5 > p_2$.

\subsection{Network Topology 3}

\begin{figure}[t]
\begin{center}
\includegraphics[scale=0.65]{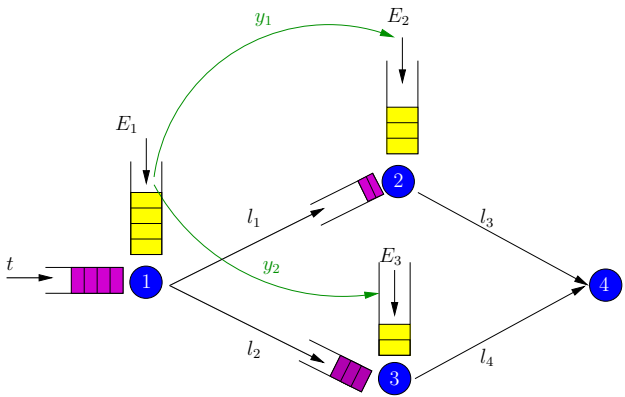}
\end{center}
\caption{Network topology 3.}
\label{sysmod4}
\end{figure}

\begin{figure}[t]
\centerline{\begin{tabular}{c}
\subfigure{\includegraphics[width=0.8\linewidth]{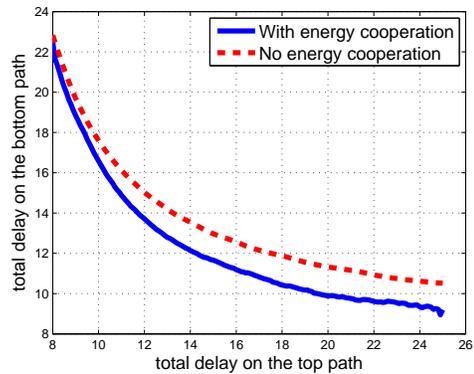}} \\
\subfigure{\includegraphics[width=0.8\linewidth]{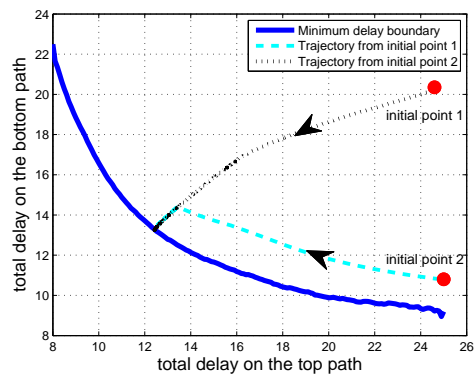}}\\
\end{tabular}}
\caption{(top) Achievable delay regions with and without energy cooperation. (bottom) Convergence of our algorithm.}
\label{delayregionpath}
\end{figure}

In this last numerical example, we demonstrate the joint optimization of flow allocation and capacity assignment. We consider the diamond network topology shown in Fig.~\ref{sysmod4} where one source is communicating with one destination with two relays in between.  The only exogenous data arrival to the network occurs at node 1 with the amount $t = 2$ units. The energy arrivals are $[E_1, E_2, E_3] = [2, 0.5, 1.5]$. Energy transfer efficiencies are given as $\alpha_1 = \alpha_2 = 0.8$. In this topology, there are six unknowns to be determined, i.e., $p_1, p_2, t_1, t_2, y_1, y_2$. By exhaustively searching over these parameters, we can obtain the minimum achievable delay region as shown in Fig.~\ref{delayregionpath}(top). In the diamond network, there are two paths of data flow. One is the top path which includes links $l_1$ and $l_3$ and the other is the bottom path which includes links $l_2$ and $l_4$. In Fig.~\ref{delayregionpath}(top), we plot the delay on bottom path versus the delay on top path. Any delay which is to the interior of this curve is achievable whereas other delays are not. All points on this boundary are Pareto-optimal points. We observe that energy cooperation enhances the achievable delay region. In Fig.~\ref{delayregionpath}(bottom), we demonstrate the convergence of our algorithm to a Pareto-optimal point. We start our algorithm from two different initial points and observe that they converge to a point which is on the boundary of the achievable delay region, demonstrating Remark~\ref{pareto}.

\section{Conclusion}

We considered the energy management and energy routing problems for delay minimization in energy harvesting networks with energy cooperation. In this network, there are data links where data flows and energy links where energy flows. We determined the jointly optimal data and energy flows in the network and the energy distribution over outgoing data links at all nodes. We established necessary conditions for the solution, and proposed an iterative algorithm that updates powers, data routing and energy routing sequentially and converges to a Pareto-optimal operating point. In the special case of fixed data flows and no energy cooperation, we showed that each link should allocate more power to links with more noise and/or more data flow. In the case with multiple energy harvests, and no energy cooperation, we showed that the optimal sum powers on the outgoing data links of each node at every slot must be equal to the optimal single-link transmit powers. Our numerical results indicate that when data flows are fixed, energy is routed from nodes with lower data loads to nodes with higher data loads; while in the more general problem, where data flows are optimized also, allocation of data and energy flows are performed in a balanced fashion.

\appendices

\section{Derivation of (\ref{optp})} \label{appendixlambert}

Starting from (\ref{optpeq}), we have
\begin{align}
\lambda = \frac{t_l}{2 \sigma_l} \left[ \frac{1}{2} \log{ \left(1 + \frac{p_l}{\sigma_l} \right)} - t_l \right]^{-2} \left( 1 + \frac{p_l}{\sigma_l} \right)^{-1} \label{kktpowre}
\end{align}
We let $r_l \triangleq \frac{1}{2} \log{ \left(1 + \frac{p_l}{\sigma_l} \right)} - t_l$, then $1 + \frac{p_l}{\sigma_l} = e^{2(r_l + t_l)}$. With these definitions, we rewrite (\ref{kktpowre}):
\begin{align}
\lambda = \frac{t_l}{2 \sigma_l} r_l^{-2} e^{-2 (r_l + t_l)}
\end{align}
Or equivalently,
\begin{align}
r_l e^{r_l} & = \sqrt{\frac{t_l  e^{-2 t_l}}{2 \lambda \sigma_l}}
\end{align}
From here, $r_l = W(z_l)$ where $z_l \triangleq \sqrt{\frac{t_l  e^{-2 t_l}}{2 \lambda \sigma_l}}$ and $W(\cdot)$ is the Lambert W function defined as the inverse function of $w \rightarrow we^w$ \cite{corless1996lambertw}. From the definition of $r_l$,
\begin{align}
\frac{1}{2} \log{ \left(1 + \frac{p_l}{\sigma_l} \right)} - t_l &= r_l = W(z_l)
\end{align}
and
\begin{align}
p_l & = \sigma_l \left( e^{2 \left( W(z_l) + t_l \right)}  -1\right)
\end{align}
which is (\ref{optp}).

\section{Derivation of (\ref{diffpower})} \label{appendixdiffpower}

From (\ref{optp}), we have
\begin{align}
p_l & = \sigma_l \left( e^{2 \left( W(z_l) + t_l \right)}  -1\right)
\end{align}
with $z_l = \sqrt{\frac{t_l  e^{-2 t_l}}{2 \lambda \sigma_l}}$. Our aim is to find $\frac{\partial p_l}{\partial \sigma_l}$. To this end, define $v_l \triangleq e^{2(W(z_l) + t_l)} -1$. Now we have,
\begin{align}
\bpar{p_l}{\sigma_l} = \left( e^{2 \left( W(z_l) + t_l \right)} - 1 \right)+ \sigma_l \bpar{v_l}{z_l} \bpar{z_l}{\sigma_l} \label{difflaw}
\end{align}
The first partial derivative on the right hand side of (\ref{difflaw}) is,
\begin{align}
\bpar{v_l}{z_l} & = e^{2 \left( W(z_l) + t_l \right)} 2 \frac{W(z_l)}{z_l(1 + W(z_l))} \label{term1}
\end{align}
where we have used (\ref{lambertprop}). The second partial derivative in (\ref{difflaw}) is,
\begin{align}
\bpar{z_l}{\sigma_l} = - \frac{1}{2} \sqrt{\frac{t_l  e^{-2 t_l}}{2 \lambda \sigma_l}} \frac{1}{\sigma_l} = - \frac{1}{2} \frac{z_l}{\sigma_l} \label{term2}
\end{align}
Using (\ref{term1}) and (\ref{term2}) in (\ref{difflaw}), we have
\begin{align}
\bpar{p_l}{\sigma_l} = e^{2 t_l} \frac{e^{2 W(z_l)}}{1 + W(z_l)} - 1
\end{align}
which is (\ref{diffpower}).

\section{Derivation of (\ref{diffflow})} \label{appendixdiffflow}

Starting from (\ref{optp}), we have
\begin{align}
p_l + \sigma_l = \sigma_l e^{2 \left( W(z_l) + t_l \right)} \label{flow1}
\end{align}
with $z_l = \sqrt{\frac{t_l  e^{-2 t_l}}{2 \lambda \sigma_l}}$. Our aim is to find $\bpar{p_l}{t_l}$. Taking logarithm of (\ref{flow1}), and differentiating both sides with respect to $t_l$, we have
\begin{align}
\frac{1}{2} \frac{1}{\sigma_l + p_l} \bpar{p_l}{t_l} &= \bpar{W(z_l)}{z_l} \bpar{z_l}{t_l} + 1  \label{flow2}
\end{align}
$\bpar{W(z_l)}{z_l}$ is evaluated from (\ref{lambertprop}) and $\bpar{z_l}{t_l}$ is
\begin{align}
\bpar{z_l}{t_l} & =  \frac{1}{2} \sqrt{\frac{t_l  e^{-2 t_l}}{2 \lambda \sigma_l}} \frac{1}{t_l} - \sqrt{\frac{t_l  e^{-2 t_l}}{2 \lambda \sigma_l}} = z_l \left( \frac{1}{2 t_l} - 1 \right) \label{flowterm2}
\end{align}
Using (\ref{lambertprop}) and (\ref{flowterm2}) in (\ref{flow2}), we obtain
\begin{align}
\bpar{p_l}{t_l} &= 2 (\sigma_l + p_l) \left[ \frac{W(z_l)}{1 + W(z_l)} \left( \frac{1}{2 t_l} - 1 \right) + 1 \right] \\
& = \sigma_l e^{2 (W(z_l) + t_l)} \frac{W(z_l) + 2 t_l }{t_l (1 + W(z_l) )}
\end{align}
which is (\ref{diffflow}).

\section{Proof of Lemma \ref{lemmasumpow}} \label{appendixsumpow}

Assume that sum powers at each slot $\bsum \triangleq \sum_{l} \bpow$ is given for each $i$. Consider the inner optimization in (\ref{problemtimenoet4}) for a fixed slot, say slot $i$. For convenience, we drop the slot index $i$, and denote $s_i$ by $s$, and $b_{li}$ by $b_l$. We define a function $x(s)$ as the minimization over $b_l$ for fixed $s$ as follows:
\begin{align}
x(s) = \min_{b_l} \qquad & \sum_{l} \frac{t_l}{\frac{1}{2} \log{\left(e^{2 t_l} + \frac{b_l}{\sigma_l}\right)} - t_l} \nonumber \\
\mbox{s.t.} \qquad & \sum_{l} b_l  = s, \quad  b_l \geq 0, \quad \forall l
\label{sumpow1}
\end{align}
which is the inner optimization in (\ref{problemtimenoet4}) for fixed $i$, and is also equivalent to:
\begin{align}
x(s) = \min_{b_l} \qquad & \sum_{l} \frac{t_l}{\frac{1}{2} \log{\left(e^{2 t_l} + \frac{b_l}{\sigma_l}\right)} - t_l} \nonumber \\
\mbox{s.t.} \qquad & \sum_{l} b_l  \leq s, \quad b_l \geq 0, \quad \forall l
\label{sumpow2}
\end{align}

Now, we claim that $x(s)$ is non-increasing and convex in $s$. Since increasing $s$ can only expand  the feasible set, $x(s)$ is non-increasing in $s$. To prove the convexity: Let $s_1, s_2 \in \mathbf{R}^+$. Let $0 \leq \lambda \leq 1$ and $\blambda = 1 - \lambda$. Let $\mathbf{b}_1$ be the solution of the problem with $s_1$, and $\mathbf{b}_2$ be the solution of the problem with $s_2$. Note that $\mathbf{b}_1$ and $\mathbf{b}_2$ exist and are unique due to convexity. The vector $\lambda \mathbf{b}_1 + \blambda \mathbf{b}_2$ is feasible for the problem with $\lambda s_1 + \blambda s_2$ since the constraints are linear. Then,
\begin{align}
x(\lambda s_1 + \blambda s_2) & \leq \sum_{l} \frac{t_l}{\frac{1}{2} \log{\left(e^{2 t_l} + \frac{\lambda b_{1l} + \blambda b_{2l}}{\sigma_l}\right)} - t_l}  \label{sumpow3} \\
& \leq \sum_l \frac{\lambda t_l}{\frac{1}{2} \log{\left(e^{2 t_l} + \frac{b_{1l}}{\sigma_l}\right)} - t_l}  \nonumber \\
& \quad + \frac{\blambda t_l}{\frac{1}{2} \log{\left(e^{2 t_l} + \frac{b_{2l}}{\sigma_l}\right)} - t_l} \label{sumpow4} \\
& = \lambda x(s_1) +  \blambda x(s_2) \label{sumpow5}
\end{align}
where (\ref{sumpow3}) follows because the minimum value of the problem can be no larger than the objective value of any feasible point, (\ref{sumpow4}) follows from the convexity of $\frac{1}{\log(a+x)}$, and (\ref{sumpow5}) follows from the fact that $\mathbf{b}_1$ solves the problem with $s_1$ and $\mathbf{b}_2$ solves the problem with $s_2$. Now, the optimization problem in (\ref{problemtimenoet4}) can be written as:
\begin{align}
\min_{\bsum} \qquad & \sum_{i=1}^T x(s_i) \nonumber \\
\mbox{s.t.} \qquad & \sum_{i=1}^k \bsum  \leq \sum_{i=1}^k G_i, \quad \forall i, \, \forall k
\label{sumpow6}
\end{align}
The problem in (\ref{sumpow6}) is in the same form as the problems in \cite[eqn.~(2)]{kayatcom12}, \cite[eqns.~(6)-(8)]{ozel11} and \cite[eqn.~(15)]{finite} and is equivalent to the problem in \cite[eqn.~(3)]{jingtcom12}, where a concave non-decreasing function of powers is maximized subject to energy harvesting constraints. In addition, \cite{kayatcom12, ozel11, finite} have additional finite battery constraints which we do not have here. References \cite{jingtcom12,kayatcom12} showed that the solution to this problem is invariant to the specific form of the function as long as it is convex (in minimization problems) or concave (in maximization problems). We follow the proof in \cite[Appendix~B]{finite} and conclude that $\bsumv$, the optimal solution of (\ref{sumpow6}), is given by the single-link optimal transmit powers.

\section{Proof of Lemma \ref{jointlemma}} \label{appendixjoint}

We show that the conditions in (\ref{ncon1})-(\ref{ncon3}) are equivalent to (\ref{jointKKT1})-(\ref{jointKKT3}) therefore proving the necessity statement of the lemma.\\
1) Writing (\ref{jointKKT1}) for node $n$ and the data links $l \in \mathcal{O}_d(n)$ connected to it
\begin{align}
\frac{t_l}{2 \sigma_l} \left[ \frac{1}{2} \log{ \left(1 + \frac{p_l}{\sigma_l} \right)} - t_l  \right]^{-2} \left( 1 + \frac{p_l}{\sigma_l} \right)^{-1} & = \lambda_{n} - \beta_l \nonumber \\ 
& \leq \lambda_{n}
\end{align}
Now, we claim that when $p_l > 0$, $\beta_l=0$. Assume $p_l > 0$ and $\beta_l > 0$. From (\ref{jointslack4}), this means that $p_l = \sigma_l (e^{2 t_l} -1)$ and the delay at link $l$ becomes $\frac{t_l}{0}$ which is unbounded for $t_l > 0$. Then, we must have $t_l = 0$, but this means $p_l = 0$, as otherwise power has been consumed on a link with zero flow. This is a contradiction to $p_l > 0$. Thus, $\beta_l = 0$ when $p_l > 0$ and (\ref{ncon1}) is satisfied with equality.\\
2) We choose any origin destination pair $(n,d)$ and identify a path starting from node $n$ and ending at destination node $d$, and in which all link powers and therefore flows are strictly positive. We denote this path by $\mathcal{F}_{n,d}$. We write the conditions (\ref{jointKKT2}) on links on this path and sum them to get
\begin{align}
\sum_{l \in \mathcal{F}_{n,d}} & \frac{1}{2} \log{ \left(1 + \frac{p_l}{\sigma_l} \right)} \left[ \frac{1}{2} \log{\left(1 + \frac{p_l}{\sigma_l} \right)} - t_l  \right]^{-2}  \nonumber \\
& = \sum_{l \in \mathcal{F}_{n,d}} \nu_{m(l)} - \nu_{n(l)} - 2 \beta_l \sigma_l e^{2 t_l} + \gamma_l \\
& = \sum_{l \in \mathcal{F}_{n,d}} \nu_{m(l)} - \nu_{n(l)} \label{jointlemmaeq1} \\
& = \nu_d - \nu_n \label{jointlemmaeq2} \\
& = - \nu_n \label{jointlemmaeq3}
\end{align}
where (\ref{jointlemmaeq1}) follows from $\beta_l = \gamma_l = 0$ since $p_l > 0$, $t_l > 0$, (\ref{jointlemmaeq2}) follows from telescoping the sum $\sum_{l \in \mathcal{F}_{n,d}} \nu_{n(l)} - \nu_{m(l)}$, and (\ref{jointlemmaeq3}) follows from setting $\nu_d = 0$ since it is a destination node and there are no flow conservation constraints at that node. We let $\tilde{\nu}_n = - \nu_n$ and get (\ref{ncon2}). \\
3) For energy link $q$ between nodes $n$ and $m$, $k(q) = n$ and $z(q) = m$ in (\ref{jointKKT3}). From (\ref{jointKKT3}), we have $\lambda_n = \alpha_q \lambda_m + \theta_q \geq \alpha_q \lambda_m$ since $\theta_q \geq 0$. Then,
\begin{align}
& \frac{t_l}{2 \sigma_l} \left[ \frac{1}{2} \log{\left( 1 + \frac{p_l}{\sigma_l} \right)} - t_l  \right]^{-2} \left( 1 + \frac{p_l}{\sigma_l} \right)^{-1} \nonumber \\
& = \lambda_n \label{jointlemmaeq4} \\
& \geq \alpha_q \lambda_m \\
& = \alpha_q \frac{t_k}{2 \sigma_k} \left[ \frac{1}{2} \log{\left(1 + \frac{p_k}{\sigma_k} \right)} - t_k  \right]^{-2} \left( 1 + \frac{p_k}{\sigma_k} \right)^{-1} \label{jointlemmaeq5}
\end{align}
where (\ref{jointlemmaeq4}) and (\ref{jointlemmaeq5}) are from using part 1 of Lemma~\ref{jointlemma} for node $n$ and $m$, respectively. Equality is achieved when $y_q > 0$, since in this case $\theta_q = 0$ from (\ref{jointslack2}).


\begin{thebibliography}{10}

\bibitem{gurakan_subm}
B.~Gurakan, O.~Ozel, J.~Yang, and S.~Ulukus, ``Energy cooperation in energy
  harvesting communications,'' {\em IEEE Trans. Comm.}, vol.~61,
  pp.~4884--4898, December 2013.

\bibitem{Bertsekas92}
D.~Bertsekas and R.~Gallager, {\em Data Networks}.
\newblock Prentice-Hall, Inc., 1992.

\bibitem{cover_book}
T.~M. Cover and J.~A. Thomas, {\em Elements of Information Theory}.
\newblock Wiley-Interscience, second~ed., 2006.

\bibitem{gallager1977minimum}
R.~G. Gallager, ``A minimum delay routing algorithm using distributed
  computation,'' {\em IEEE Trans. Comm.}, vol.~25, pp.~73--85, January 1977.

\bibitem{bertsekas1984second}
D.~P. Bertsekas, E.~M. Gafni, and R.~G. Gallager, ``Second derivative
  algorithms for minimum delay distributed routing in networks,'' {\em IEEE
  Trans. Comm.}, vol.~32, pp.~911--919, August 1984.

\bibitem{gavish1989system}
B.~Gavish and I.~Neuman, ``A system for routing and capacity assignment in
  computer communication networks,'' {\em IEEE Trans. Comm.}, vol.~37,
  pp.~360--366, April 1989.

\bibitem{bertsekas1998network}
D.~Bertsekas, {\em Network Optimization: Continuous and Discrete Models}.
\newblock Athena Scientific, 1998.

\bibitem{yen2001near}
H.~H. Yen and F.~S. Lin, ``Near-optimal delay constrained routing in virtual
  circuit networks,'' in {\em IEEE INFOCOM}, April 2001.

\bibitem{cruz2003optimal}
R.~L. Cruz and A.~V. Santhanam, ``Optimal routing, link scheduling and power
  control in multihop wireless networks,'' in {\em IEEE INFOCOM}, March 2003.

\bibitem{xiao2004simultaneous}
L.~Xiao, M.~Johansson, and S.~Boyd, ``Simultaneous routing and resource
  allocation via dual decomposition,'' {\em IEEE Trans. Comm.}, vol.~52,
  pp.~1136--1144, July 2004.

\bibitem{cui2007cross}
S.~Cui, R.~Madan, A.~J. Goldsmith, and S.~Lall, ``Cross-layer energy and delay
  optimization in small-scale sensor networks,'' {\em IEEE Trans. Wireless
  Comm.}, vol.~6, pp.~3688--3699, October 2007.

\bibitem{xi2008node}
Y.~Xi and E.~M. Yeh, ``Node-based optimal power control, routing, and
  congestion control in wireless networks,'' {\em IEEE Trans. Inform. Theory},
  vol.~54, pp.~4081--4106, September 2008.

\bibitem{jingtcom12}
J.~Yang and S.~Ulukus, ``Optimal packet scheduling in an energy harvesting
  communication system,'' {\em IEEE Trans. Comm.}, vol.~60, pp.~220--230,
  January 2012.

\bibitem{kayatcom12}
K.~Tutuncuoglu and A.~Yener, ``Optimum transmission policies for battery
  limited energy harvesting nodes,'' {\em IEEE Trans. Wireless Comm.}, vol.~11,
  pp.~1180--1189, March 2012.

\bibitem{ozel11}
O.~Ozel, K.~Tutuncuoglu, J.~Yang, S.~Ulukus, and A.~Yener, ``Transmission with
  energy harvesting nodes in fading wireless channels: Optimal policies,'' {\em
  IEEE JSAC}, vol.~29, pp.~1732--1743, September 2011.

\bibitem{ho-zhang-tsp12}
C.~K. Ho and R.~Zhang, ``Optimal energy allocation for wireless communications
  with energy harvesting constraints,'' {\em IEEE Trans. Signal Proc.},
  vol.~60, pp.~4808--4818, September 2012.

\bibitem{finite}
O.~Ozel, J.~Yang, and S.~Ulukus, ``Optimal broadcast scheduling for an energy
  harvesting rechargeable transmitter with a finite capacity battery,'' {\em
  IEEE Trans. Wireless Comm.}, vol.~11, pp.~2193--2203, June 2012.

\bibitem{kaya13ita}
K.~Tutuncuoglu and A.~Yener, ``Multiple access and two-way channels with energy
  harvesting and bidirectional energy cooperation,'' in {\em UCSD ITA
  Workshop}, February 2013.

\bibitem{kaya13itw}
K.~Tutuncuoglu and A.~Yener, ``Cooperative energy harvesting communications
  with relaying and energy sharing,'' in {\em IEEE ITW}, September 2013.

\bibitem{varshney2008}
L.~R. Varshney, ``Transporting information and energy simultaneously,'' in {\em
  IEEE ISIT}, July 2008.

\bibitem{grover2010}
P.~Grover and A.~Sahai, ``{S}hannon meets {T}esla: wireless information and
  power transfer,'' in {\em IEEE ISIT}, July 2010.

\bibitem{doost}
R.~Doost, K.~R. Chowdhury, and M.~{Di Felice}, ``Routing and link layer
  protocol design for sensor networks with wireless energy transfer,'' in {\em
  IEEE Globecom}, December 2010.

\bibitem{lshi}
L.~Xie, Y.~Shi, {Y. T. Hou}, W.~Lou, H.~Sherali, and S.~Midkiff, ``On renewable
  sensor networks with wireless energy transfer: {T}he multi-node case,'' in
  {\em IEEE SECON}, June 2012.

\bibitem{zhang2013mimo}
R.~Zhang and C.~K. Ho, ``{MIMO} broadcasting for simultaneous wireless
  information and power transfer,'' {\em IEEE Trans. Wireless Comm.}, vol.~12,
  pp.~1989--2001, May 2013.

\bibitem{zhou2012wireless}
X.~Zhou, R.~Zhang, and C.~K. Ho, ``Wireless information and power transfer:
  Architecture design and rate-energy tradeoff,'' {\em IEEE Trans. Comm.},
  vol.~61, pp.~4754--4767, November 2013.

\bibitem{simeonegraph}
A.~Fouladgar and O.~Simeone, ``Information and energy flows in graphical
  networks with energy transfer and reuse,'' {\em IEEE Wireless Comm. Letters},
  vol.~2, pp.~371--374, August 2013.

\bibitem{huang2014}
K.~Huang and V.~Lau, ``Enabling wireless power transfer in cellular networks:
  Architecture, modeling and deployment,'' {\em IEEE Trans. Wireless Comm.},
  vol.~13, pp.~902--912, February 2014.

\bibitem{gurakanisit2014}
B.~Gurakan and S.~Ulukus, ``Energy harvesting diamond channel with energy
  cooperation,'' in {\em IEEE ISIT}, June 2014.

\bibitem{ding2015application}
Z.~Ding, C.~Zhong, D.~W.~K. Ng, M.~Peng, H.~Suraweera, R.~Schober, and H.~Poor,
  ``Application of smart antenna technologies in simultaneous wireless
  information and power transfer,'' {\em IEEE Comm. Mag.}, vol.~53, no.~4,
  pp.~86--93, 2015.

\bibitem{hu2013utility}
C.~Hu, X.~Zhang, S.~Zhou, and Z.~Niu, ``Utility optimal scheduling in energy
  cooperation networks powered by renewable energy,'' in {\em IEEE APCC},
  pp.~403--408, 2013.

\bibitem{guo2014joint}
Y.~Guo, J.~Xu, L.~Duan, and R.~Zhang, ``Joint energy and spectrum cooperation
  for cellular communication systems,'' {\em IEEE Trans. Comm.}, vol.~62,
  no.~10, pp.~3678--3691, 2014.

\bibitem{leithon2014energy}
J.~Leithon, T.~Lim, and S.~Sun, ``Energy exchange among base stations in a
  cellular network through the smart grid,'' in {\em IEEE ICC}, pp.~4036--4041,
  2014.

\bibitem{chia2014energy}
Y.~K. Chia, S.~Sun, and R.~Zhang, ``Energy cooperation in cellular networks
  with renewable powered base stations,'' {\em IEEE Trans. Wireless Comm.},
  vol.~13, no.~12, pp.~6996--7010, 2014.

\bibitem{xu2015cost}
J.~Xu, L.~Duan, and R.~Zhang, ``Cost-aware green cellular networks with energy
  and communication cooperation,'' {\em IEEE Comm. Mag.}, vol.~53, no.~5,
  pp.~257--263, 2015.

\bibitem{goldsmith-varaiya}
A.~J. Goldsmith and P.~P. Varaiya, ``Capacity of fading channels with channel
  side information,'' {\em IEEE Trans. on Inform. Theory}, vol.~43,
  pp.~1986--1992, November 1997.

\bibitem{boyd}
S.~Boyd and L.~Vandenberghe, {\em Convex Optimization}.
\newblock Cambridge University Press, 2004.

\bibitem{corless1996lambertw}
R.~M. Corless, G.~H. Gonnet, D.~E. Hare, D.~J. Jeffrey, and D.~E. Knuth, ``On
  the {L}ambert {W} function,'' {\em Adv. in Comp. Math.}, vol.~5,
  pp.~329--359, December 1996.

\end{thebibliography}
\end{document}